\documentclass[%
 aip,
 sd,%
 amsmath,amssymb,
 reprint,%
]{revtex4-1}

\usepackage{multirow}
\usepackage{graphicx}
\usepackage{bm}% bold math
\usepackage{hyperref}% add hypertext capabilities
\usepackage{graphicx, subfigure, textcomp, amssymb, mathrsfs}

\usepackage[version=3]{mhchem} 
\usepackage{chemarr}

\usepackage{epstopdf}
\usepackage{soul,color}

\begin{document}

\title{Modified Poisson-Nernst-Planck theory for low-to-mid frequency immittance of electric double-layer capacitors}

\author{Anis Allagui$^*$}
\email{aallagui@sharjah.ac.ae}
\affiliation{Dept. of Sustainable and Renewable Energy Engineering, University of Sharjah, Sharjah, P.O. Box 27272, United Arab Emirates}
\altaffiliation[Also at ]{Center for Advanced Materials Research, Research Institute of Sciences and Engineering, University of Sharjah, Sharjah,, P.O. Box 27272,  United Arab Emirates}
\affiliation{Dept. of Mechanical and Materials Engineering, Florida International University, Miami, FL33174, United States}

\author{Hachemi Benaoum} 
%\email{hbenaoum@sharjah.ac.ae}
\affiliation{
Dept. of Applied Physics and Astronomy, 
University of Sharjah, PO Box 27272, Sharjah, United Arab Emirates 
}

\author{Hichem Eleuch} 
%\email{hbenaoum@sharjah.ac.ae}
\affiliation{
Dept. of Applied Physics and Astronomy, 
University of Sharjah, PO Box 27272, Sharjah, United Arab Emirates 
}

\author{Chunlei Wang} 
\affiliation{Dept. of Mechanical and Materials Engineering, Florida International University, Miami, FL33174, United States}

\begin{abstract}

Understanding the system-level spectral immittance response of capacitive energy storage devices with analytically tractable physics-based models is not only important for the progress of the technology, but also allows to develop new physical insights more easily. Here, we report a modified Poisson–Nernst–Planck (PNP) system  describing charge concentration and electric potential as a model of electrokinetics for electrodes showing mixed resistive-capacitive behavior. This is done by (i) incorporating time shifts between the current fluxes and both concentration gradients of charged species and the electric field, and (ii) introducing time fractional derivatives in the continuity equation. The aim is to characterize  the deviation of immittance from that of ideal capacitors both at close-to-dc frequencies where the impedance angle for example is larger than -90 deg., and also at mid-range frequencies where the system veers progressively toward resistive behavior. This latter tendency is important to model  in order to identify the extend of the capacitive  bandwidth of the device from the rest. Solution and simulation results to the one-dimensional modified PNP system for symmetric electrolyte/blocking electrode configuration are presented and discussed.  

\end{abstract}

\maketitle

\section{Introduction}
\label{introduction} 
  
  The analysis and interpretation of measured spectral immittance profile  is commonly carried out using equivalent circuit models.  In particular, the  non-ideal behavior that cannot be properly captured by circuits of $R$, $L$ and $C$ elements is  usually  described by circuits containing the constant phase element (CPE) \cite{foedlc,EC2015,fracorderreview,eis,cpe}. 
    The CPE, also known as  mono-order fractional  capacitor, is widely employed for the modeling of electric double-layer capacitors (EDLC) and porous electrodes \cite{bisquert1998impedance, EC2015, fracorderreview, ieeeted, allagui2021inverse, memQ, memoryAPL, JMCA, ACSApplEnergyMater}, batteries \cite{LIB1, cheng2018time, xiong2018novel, cuervo2015unifying}, solar cells \cite{orgElectronics,oe2}, and sensor devices \cite{borini2013ultrafast, song2020enhanced, meunier2020interpreting}.  
It has the (empirical)  power-law complex impedance function of the form:
\begin{equation}
Z(s) = V(s)/I(s) = 1/(s^{\alpha} C_{\alpha}) 
\label{eq:1}
\end{equation}
  where $C_{\alpha}$ is   in units of F\,s$^{\alpha-1}$ (pseudocapacitance), $0<\alpha <1$ and $s=j\omega$ the complex angular frequency.  The real and imaginary parts of a CPE are 
 $\mathrm{Re}(Z) = \omega^{-\alpha} C_{\alpha}^{-1} \cos(\alpha\pi/2) $, and  $\mathrm{Im}(Z) =- \omega^{-\alpha} C_{\alpha}^{-1} \sin(\alpha\pi/2) $, respectively, and thus a constant, frequency-independent impedance phase angle $\phi(Z) = -\alpha \pi/2$, which can take values anywhere between -90 deg. and zero. 
 We note that in the  time-domain, the CPE is characterized by the current-voltage fractional order  differential equation \cite{fracorderreview,EC2015, memQ,allagui2021possibility}:  
 \begin{equation}
i(t) = C_{\alpha}\, _0D_t^{\alpha} v(t)  
\label{eqRL}
 \end{equation}
 where  $_0D_t^{\alpha} v(t)$ is defined here in the  Caputo sense  by \cite{podlubny1998fractional}:     
 \begin{equation}
 _0D_t^{\alpha} v(t) =  \frac{1}{\Gamma(1-\alpha)} \int_0^t   \frac{\mathrm{d}v(\tau)/\mathrm{d}\tau}{(t-\tau)^{\alpha}}      d\tau
%= \frac{1}{\Gamma(1-\alpha)} \frac{d}{dt} \int\limits_0^t v(x) (t-x)^{-\alpha} dx
\label{eqRL}
\end{equation}    
For the limiting case of $\alpha = 0$, the impedance in Eq.\;\ref{eq:1} is real-valued, $Z= R$ 
(i.e. $1/C_{\alpha}  \leftarrow R$) for any frequency, and Eq.\;\ref{eqRL} turns to be Ohm's law $i(t)=v(t)/R$ (zero-order derivative   $_0{D}_t^{0} f(t)$ gives $f(t)$, i.e. the identity operator). 
 Whereas for $\alpha=1$, the impedance is that of an ideal capacitor $1/(sC_1)$ and the current-voltage characteristic equation is $i(t)=C_1 \mathrm{d}v(t)/\mathrm{d}t$. The Warburg impedance is in fact a special case of the CPE corresponding to $\alpha=1/2$. 
  If the value of $\alpha$   is taken from  $ \left]-1;0\right[$, we may speak of fractional inductors of mixed behavior lying    between that of  ideal inductors and resistors \cite{energy2015}, but that is not the focus of this work.  

\begin{figure*}[t]
\begin{center}
\subfigure[]{\includegraphics[width=2.5in]{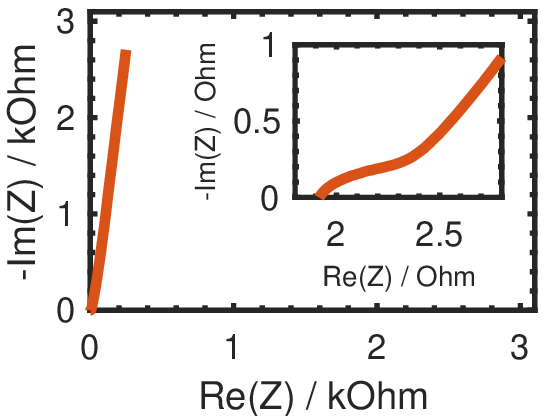}}
\subfigure[]{\includegraphics[width=2.5in]{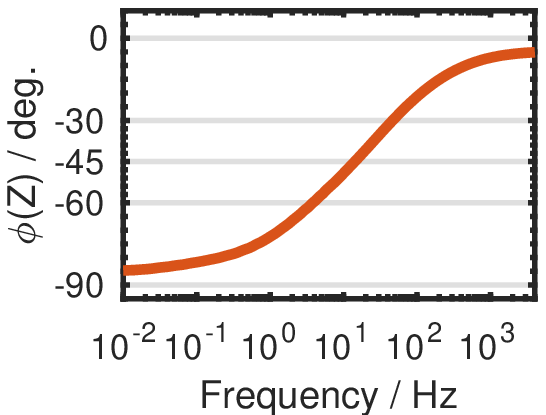}}
\caption{Results of open-circuit electrochemical impedance spectroscopy of an all-solid-state (symmetric)  multi-layer graphene electric double-layer capacitor \cite{bpf,2017-3}: (a) Nyquist plot of imaginary part vs. real part (frequency is swept as a parameter), and (b) impedance phase angle vs. log of frequency.}
\label{fig0}
\end{center}
\end{figure*}

Because of the extra degree of freedom in the CPE impedance, that is the dispersion coefficient $\alpha$, this element is proved successful for describing phenomenologically   the inclined and depressed   Nyquist impedance plots of real vs. imaginary parts of many systems, as well as their time-domain responses \cite{eis,LIB1,cpe, EC2015, energy2015, orgElectronics}. 
In fact the pure capacitive behavior is rarely encountered in real  measurements which made the CPE a very successful fitting tool. 
However, 
 typical experimental results,  as shown in  Fig.\;\ref{fig0} for  multi-layered graphene-based   EDLC, exhibit fractional order behavior  in the close-to-dc region   (impedance phase angle is practically constant and equal to -83$^{\circ}$ over the frequency bandwidth 0.01 to $<1$ Hz), but as the frequency is increased we observe an increase of resistive behavior at the expense of the capacitive behavior. The low-frequency segment can  be fitted well enough with a single CPE or a CPE with a resistor, but that cannot be the case anymore as the frequency is increased. It is  usually required to have additional elements in the modeling circuit which can complicate the analysis and hinder the physical tractability to the system under consideraion.  
Furthermore, the CPE electrical parameters have still unsatisfactory  interpretations, and their relationship   to real physical measurables is still  a subject of debate. The current understanding   revolves  around subdiffusive charge transport where the mean square displacement of a particle  follows a power law function of time \cite{kosztolowicz2009hyperbolic, foedlc}, fractal morphologies and interfacial roughness of electrodes \cite{kant2008theory, maritan1989anomalous}, or some form of distribution of response relaxation \cite{zhang2015reconstruction}.

 The motivation behind this work is to study the anomalous frequency-domain  immittance function observed with EDLCs at a system level using physics-based models. 
 Our goal is to   describe both the  deviation of the frequency response   of such devices from that of  ideal capacitors at low frequencies (i.e. CPE behavior), but also to show the evolution of the system towards resistive behavior as the frequency is increased. 
To this end we employ the standard electrokinetic model based on Poisson-Nernst-Planck  (PNP) equations, which  is well-known   for treating electrochemical transport of charged species   by  relating their fluxes   to  both  concentration gradients and the electric field \cite{barbero2017theoretical}. 
 It has been applied and validated in electrochemistry and analytical chemistry,  sensor devices, and in biology \cite{kilic2007steric, schmuck2015homogenization, alexe2020current, koklu2019self, barbero2017theoretical}. In addition, these equations also appear in plasma physics, and semiconductor device modeling, where they are known as the diffusion-drift equations or the van Roosbroeck equations. 
 We   consider  the case of perfect blocking electrodes
 without   convective electrolyte transport, steric effects due to finite ion sizes,  or Faradaic electrochemical reactions at the electrodes.  
 We will, however, incorporate  a time shift between the fluxes of charged species and both their concentration gradients  and the electric field in a similar way as Cattaneo's constitutive formulation \cite{kosztolowicz2009hyperbolic, compte1997generalized}. 
We recall that in Cattaneo's  formulation of Fickian diffusion for instance, the constitutive equation relating flux $\mathbf{J}(\mathbf{x},t)$ to the distribution function of the diffusing species $c(\mathbf{x},t)$, i.e. 
 \begin{equation}
\mathbf{J}(\mathbf{x},t)=-D \,\nabla c(\mathbf{x},t)
\label{eq04}
\end{equation}
 would be considered (up to the first order)  as: 
 \begin{equation}
\mathbf{J}(\mathbf{x},t) + \tau_c \frac{\partial \mathbf{J}(\mathbf{x},t)}{\partial t}=-D\, \nabla \mathbf{c}(\mathbf{x},t)
\label{eq06}
\end{equation} 
with non-zero value of the relaxation time constant $\tau_c$ (typically in the range of $10  \sim 100 $~s for porous materials and biological tissues,  $10^{-10}  \sim 10^{-14} $ s for metals, $10^{-8}  \sim 10^{-10}$~s for gases). 
 The phenomenological derivation of this equation is based on the assumption that the flux of  particles is not generated by the concentration gradient instantaneously at time $t$, as it is the case with the  normal diffusion equation, but   is delayed by a certain time lag $\tau_c$ \cite{lewandowska2008application}. Such an approach has been   extensively studied and verified in heat transfer (Fourier's law) \cite{ruggeri1990shock, RevModPhys.61.41} and diffusion problems (Fick's law) \cite{compte1997generalized,kosztolowicz2014cattaneo,kosztolowicz2009hyperbolic, garcia1995hyperbolic, qi2011solutions}.  
 Without this modification, the fundamental solution to the   (parabolic) diffusion equation (obtained by substituting Eq.\;\ref{eq04} in the continuity equation, ${\partial c_i (\mathbf{x},t)}/{\partial t} + \nabla  \mathbf{J}_i (\mathbf{x},t) = 0
$, Eq.\;\ref{eq4} below), i.e.    
\begin{equation}
\frac{\partial c(\mathbf{x},t)}{\partial t} = D \nabla^2 c(\mathbf{x},t)
\end{equation}
 in response to an initial Dirac delta function $c(\mathbf{x},t=0)=\delta(\mathbf{x})$,   is given by the Gaussian distribution:
 \begin{equation}
c(\mathbf{x},t)=(4\pi D t)^{-d/2} e^{-\textbf{x}^2/(4Dt)}
\label{eq7}
\end{equation}
Eq.\;\ref{eq7} clearly leads to  non-zero, finite amount of diffusing species at large distance away from the origin even for very small values of $t$ \cite{compte1997generalized}. This can be interpreted as a propagation of species taking place at infinite velocity, which is unphysical. 
 Note that  combining Eq.\;\ref{eq06}  with the unchanged continuity equation ${\partial c(\mathbf{x},t)}/{\partial t}  = -D \nabla  \mathbf{J}(\mathbf{x},t)$, one obtains the second-order hyperbolic differential equation  of diffusion as per Cattaneo's formulation \cite{compte1997generalized, kosztolowicz2009hyperbolic}:
\begin{equation}
\frac{\partial c(\mathbf{x},t)}{\partial t}  + \tau_c \frac{\partial^2 c(\mathbf{x},t)}{\partial t^2}  = D \nabla^2 c(\mathbf{x},t) 
\label{eqC} 
\end{equation} 
 In the PNP system we propose below for describing the frequency behavior of EDLCs, we will adopt a similar approach 
   which will provide, as we shall show,  the tools for capturing the tendency of the electrochemical capacitor towards resistive behavior as the frequency is increased.  
  Furthermore, we consider fractional-order time derivative in the continuity, which is commonly used for describing the anomalous behavior of complex system when integer-order models fails, such as in disordered systems, gels and porous electrodes   \cite{kosztolowicz2009hyperbolic, gomez, LENZI2021105907, qi2011solutions}. Such a generalization is connected to Hilfer's \textit{fractional invariance} or \textit{fractional stationarity} \cite{hilfer1995foundations}.

The remainder of the article is organized as follows. In Section\;\ref{theory},  we describe the classical ion transport in an  electrochemical system based on PNP theory, and the associated boundary conditions we considered to obtain an expression for the admittance/impedance functions of   ideal capacitors. We also derive the new expressions for the admittance/impedance functions (Eq.\;\ref{eq32}) by solving the modified PNP system. 
 Simulations and      
 discussion   of the results are presented  in Section\;\ref{simulation}.  Finally, in Section\;\ref{conclusion}, we present our concluding remarks highlighting  the main advantages of the modified PNP model for describing anomalous impedance of fractional EDLCs over extended frequency bandwidths. 
 
\section{Theory}
\label{theory}

\subsection{Classical and modified PNP systems}

First, we present the basic equations used for describing electrified electrodes in contact with electrolytes, which   are the flux equations, the continuity equation and the local electroneutrality assumption \cite{manzanares2003diffusion}. Assuming constant pressure and temperature, the flux equation for a given species $i$ (of positive or negative charge) in the electrolyte is derived from the difference in electrochemical potentials $\mu'_i - \mu''_i$ between two spatially adjacent regions of volume elements that we denote by $V'$ and $V''$. This is related to the change in internal energy of these two volumes which are $\text{d}U' = T\text{d}S'+ \sum_i \mu_i \text{d}n'_i $ and 
 $\text{d}U'' = T\text{d}S''+ \sum_i \mu_i \text{d}n''_i 
$, with the additional  assumption of  transport without energy exchange with the surroundings, i.e., $\text{d}U' + \text{d}U''=0$ and $\text{d}n'_i + \text{d}n''_i=0$. 
Thus, from the Second Law of Thermodynamics for the control volume $V' \cup V''$, we have $ T(\text{d}S'+\text{d}S'') = \sum_i( \mu''_i - \mu'_i) \text{d}n'_i \geqslant 0 $, which indicates that $\text{d}n'_i$ is determined only by the difference $\mu''_i - \mu'_i$ when the transport of species $i$ is not
coupled to that of other species. If we denote by $\mathbf{n}\, \text{d}A$   the area separating the two volumes and oriented along the unit vector $\mathbf{n}$ from $V'$ and $V''$, then the number of moles of species $i$ crossing the surface in a time $\text{d}t$ is $\text{d}n'_i = \ \mathbf{J}_i \, \mathbf{n}\, \text{d}A\, \text{d}t$. When the difference in electrochemical potentials is not too large,  we can assume that $\text{d}n'_i/\text{d}t$ is proportional to the gradient of the electrochemical potential normal to the surface (i.e. $\nabla \mu_i\, \mathbf{n}$), and thus express the vector flux density as: 
\begin{equation}
\mathbf{J}_i (\mathbf{x},t)= c_i (\mathbf{x},t) \mathbf{v}_i (\mathbf{x},t) = -\frac{D_i c_i (\mathbf{x},t)}{RT} \nabla \mu_i (\mathbf{x},t)
\end{equation}
Here $\mathbf{v}_i (\mathbf{x},t)= -u_i \nabla \mu_i (\mathbf{x},t)$ is the velocity of species $i$ of composition  $c_i$, where $u_i=D_i/RT$ (Einstein relation) is its mobility and $D_i$ its diffusion coefficient (thermal motion). Furthermore, $\nabla \mu_i (\mathbf{x},t)$ can be related to the changes in composition $c_i (\mathbf{x},t)$ and  in electric potential $\phi (\mathbf{x},t)$ such that:
\begin{equation}
\nabla \mu_i (\mathbf{x},t) = RT \nabla \ln c_i (\mathbf{x},t)   + z_i F \nabla \phi (\mathbf{x},t)  
\end{equation}
where $F$ is the Faraday constant and $z_i$ is the charge number of species $i$. This leads to the standard  Nernst-Plank ionic flux equation of diffusion and migration in  the domain (without convective contribution or Faradic charge transfer) being:
\begin{equation}
\mathbf{J}_i (\mathbf{x},t) =- D_i \left( \nabla  c_i (\mathbf{x},t) +  \frac{z_i F}{RT} c_i \nabla \phi (\mathbf{x},t) \right)
\label{eq3}
\end{equation}
To complete the necessary system of equations for describing the electrochemical system, the time-dependence of composition is given by the divergence of the flux (continuity equation) such that:
\begin{equation}
\frac{\partial c_i (\mathbf{x},t)}{\partial t} + \nabla  \mathbf{J}_i (\mathbf{x},t) = 0
\label{eq4}
\end{equation}
with the assumption that no chemical reactions can take place.  In addition, the electrostatic potential distribution in the domain can be determined by Poisson's equation as:
\begin{equation}
\nabla^2 \phi (\mathbf{x},t) = -\frac{\rho (\mathbf{x},t)}{\epsilon} = -\frac{F}{\epsilon} \sum_i z_i c_i (\mathbf{x},t)
\label{eq5}
\end{equation}
where $\rho (\mathbf{x},t)  = \sum_i z_i F c_{i} (\mathbf{x},t)$ is the volume charge density, and $\epsilon$ is the dielectric permittivity of the bulk solution which is assumed to be constant and frequency-independent.  Here the electric field is considered conservative and related to the potential via $\mathbf{E} (\mathbf{x},t)=-\nabla \phi (\mathbf{x},t)$.

At this point we introduce  the modified PNP system that we propose taking into account Cattaneo's approach.  
 In order to consider a finite velocity of propagation of diffusing species  and relax the flux, Cattaneo proposed to replace the standard linear response  (Eq.\;\ref{eq04}) by  Eq.\;\ref{eq06} to obtain Eq.\;\ref{eqC}.   
However,   the left-hand side of Eq.\;\ref{eq06} is nothing but the first-order approximation of the delayed flux  $\mathbf{J} (\mathbf{x},t+\tau_c)$ in the limit of small values of $\tau_c$, i.e. \cite{kosztolowicz2009hyperbolic, lewandowska2008application}: 
\begin{eqnarray}
\mathbf{J} (\mathbf{x},t+\tau_c)
&=& e^{\tau_c \frac{\partial}{\partial t}} \bold{J}_i \left(\bold{x}, t \right) \nonumber \\
& \approx& \mathbf{J} (\mathbf{x},t) + \tau_c \frac{\partial \mathbf{J} (\mathbf{x},t)}{\partial t}  + \ldots 
%= - D \nabla c (\mathbf{x},t) 
\label{eqC0}
\end{eqnarray}
from which we write:
\begin{equation}
\mathbf{J}_{i_{\alpha}}(\mathbf{x},t+\tau_c) = -\frac{{D_{i_\alpha}} c_{i}(\mathbf{x},t)}{RT}
 \nabla 
    \mu_{i}(\mathbf{x},t)
          \label{eqJa}
\end{equation}  
or
\begin{equation}
\mathbf{J}_{i_{\alpha}}(\mathbf{x},t) = -\frac{{D_{i_\alpha}} c_{i}(\mathbf{x},t-\tau_c)}{RT}
 \nabla 
    \mu_{i}(\mathbf{x},t-\tau_c)
      \label{eqJa1}
\end{equation}
Combined with the fractional continuity equation  \cite{compte1997generalized, hilfer1995foundations, qi2011solutions, hilfer2000fractional, kosztolowicz2009hyperbolic}, 
\begin{equation}
_0D_t^{\alpha} c_{i}{(\mathbf{x}, t )}  
=  
-\nabla \mathbf{J}_{i_{\alpha}}(\mathbf{x},t) \quad (0<\alpha<1)
\label{eq23} 
\end{equation}  
 the proposed modified PNP system is written   in terms of a single   lag $ \tau_c$ and fractional-order time derivative as follows:
 \begin{widetext}
 \begin{equation}
_0D_t^{\alpha} c_{i}{(\mathbf{x}, t )}  
= -\nabla \left\{
 \frac{{D_{i_\alpha}} c_{i}(\mathbf{x},t-\tau_c)}{RT}
 \nabla \left\{
   RT \ln c_{i}(\mathbf{x},t-\tau_c)
    + {z_i F }  
        \phi (\mathbf{x},t-\tau_c) 
      \right\} \right\}
\end{equation}
\end{widetext} 
or in terms of two phase lags $\tau_1$ and $\tau_2$ (for mathematical convenience, see Eq.\;\ref{eq250} below) as \cite{tzou1995unified, compte1997generalized}: 
\begin{widetext} 
\begin{equation}
_0D_t^{\alpha} c_{i}{(\mathbf{x}, t+\tau_1)}  
= -\nabla \left\{
 \frac{{D_{i_\alpha}} c_{i}(\mathbf{x},t +\tau_2)}{RT}
 \nabla \left\{
   RT \ln c_{i}(\mathbf{x},t+\tau_2)
    + {z_i F }  
        \phi (\mathbf{x},t+\tau_2) 
      \right\} \right\}
\label{eq231} 
\end{equation}
\end{widetext}
 together with the (unchanged) Poisson equation: 
\begin{equation}
\nabla^2  \phi (\mathbf{x},t)  = -\frac{\rho (\mathbf{x},t)}{\epsilon}
\label{mP}
\end{equation} 
The generalization of the continuity equation, that now includes fractional derivative of constant order $\alpha \neq 1$, stands in direct connection with Hilfer's \textit{fractional invariance} or \textit{fractional stationarity} \cite{hilfer1995foundations}.  It has been also adopted in various forms, for instance by Compte and Metzler \cite{compte1997generalized} or G{\'o}rska \cite{gorska2021integral} for the study of anomalous diffusive transport within the framework of Cattaneo, and similarly by Kosztołowicz and Lewandowska \cite{kosztolowicz2009hyperbolic} and others \cite{lenzi2009fractional} for study of subdiffusive impedance spectra, by Kosztołowicz \cite{kosztolowicz2014cattaneo, kosztolowicz2019model} for the study of  Cattaneo-type subdiffusion-reaction and anomalous diffusion-absorption processes.  
    Note that in order to maintain proper dimensions of both sides of   Eq.\;\ref{eq231},  the modified  diffusion coefficients $D_{i_\alpha}$ are in units of m$^2$/s$^{\alpha}$.  
Furthermore, it  is assumed with  the suggested model that there is delayed flux-force relation between  the  current flux on the one hand and  changes in concentration  of species $i$ and  in the electric potential in the other hand, which is taken into account via the net time shift $\Delta\tau = \tau_c = \tau_2-\tau_1$.  This can be particularly  justified for the case of spectral impedance of electrochemical systems,  especially at high frequencies at which the concentrations of species are expected to oscillate very rapidly \cite{lewandowska2008application}. Excitation and response cannot happen exactly at the same time. More details on the time shift $\Delta \tau$ will be discussed in Section.\;\ref{simulation}.  Furthermore, it is clear   from Catteneo's formalism (Eq.\;\ref{eq06})  that the delayed concentration is considered up to   the first order only, whereas in our generalized formulation we   take into account all orders without approximation.   Eq.\;\ref{eq231} simplifies to Cattaneo's-like formulation when taking first order only in terms of $\tau_1$ and setting $\tau_2=0$.    

\subsection{Modified immittance function}
 
 Recall that the goal is to derive an expression for the immittance function of the electrochemical system which necessitates an expression for the Laplace transform of the current density in response to a  voltage excitation across the electrodes. 
 In practice, an electrical stimulus (voltage or current) is applied and the response (current or voltage) is measured for a discrete set of frequencies, from which several information about the system can be determined \cite{macdonald2005impedance}.
 The system is   characterized by its transfer function
$H(s)= {\mathcal{L}{(v(t))}}/{\mathcal{L}(i(t))}$     or its reciprocal. When a steady-state sinusoidal  excitation is applied, 
 and  if the system under test is also linear and the conditions of causality and time-invariance  are obeyed (Kramers--Kronig relationships), the transfer function may be identified as the impedance $Z(j\omega)$ of the system, and its reciprocal is the admittance function $Y(j\omega)$. 
 
 For  analytical  convenience,  we    consider here the simple case of   ideally polarized electrode submerged in dilute symmetric  electrolyte with the ionic charge of positive and negative ions being $|z_{_+}|= |z_{_-}| = z$ and their diffusion coefficients being $D_{{+_{\alpha}}} = D_{-_{\alpha}}= D_{\alpha}$. We also restrict our analysis to  one-dimensional case. 
 In this case, the density of charge for any time $t$ and location $ {x}$ is   
$\rho( {x},t) = zF ( c_{_+}( {x},t) - c_{_-}( {x},t) )$. Also, in view of small perturbations of potential and charge, the sum of compositions 
 $c_{_+}( {x},t) + c_{_-}( {x},t)$ can be assumed to be constant and always equal to $2c_0$.  
 Thus, the current density  at any cross section of the system and at any time is obtained from Eq.\;\ref{eqJa}
 (or equivalently Eq.\;\ref{eq3} for the classical case) as \cite{BUCK1969219, bazant2004diffuse}:
\begin{equation}
i_{\alpha}(x,t) = -D_{\alpha} \frac{\partial \rho (x,t)}{\partial x} - D_{\alpha}   \kappa^2 \epsilon \frac{\partial \phi (x,t)}{\partial x} 
\label{eq8}
\end{equation} 
 where $\kappa$ is the reciprocal Debye thickness given by:
 \begin{equation}
\kappa =   \sqrt{2c_0 z^2 F^2/(\epsilon RT)}
\end{equation} 
Eq.\;\ref{eq8} is the sum of two currents: a conduction current  caused by concentration gradients and a displacement current  caused by the electric field. Note that although we allow the potential and currents to be time-varying, magnetic effects can be  considered  negligible in our situation \cite{bonnefont2001analysis}.

 The domain's boundary conditions we consider are such that the fluxes 
$ {J}_{+_{\alpha}}= {J}_{-_{\alpha}}=0$ at both electrodes for any time $t>0$ (i.e. no Faradaic current, perfect blocking electrodes),  and 
$\rho( {x},0) = z F \left(  c_{_+}( {x},0) -  c_{_-}( {x},0) \right) = \rho_0$ and 
$\rho( {x},\infty) = \rho_{\infty}$ \cite{BUCK1969219}.

To solve this problem we proceed as follows. 
We assume for simplicity 
  the same values of delays $\tau_1$ and $\tau_2$, and the same order of fractional derivative (i.e. $\alpha$) in Eq.\;\ref{eq231}  for describing the dynamics of both positively and negatively-charged species. 
We write the fractional time derivative of $\rho( {x},t) = zF ( c_{_+}( {x},t) - c_{_-}( {x},t) )$ with the use of 
 Eqs.\;\ref{eq231} and \ref{eqJa}  
 (or equivalently the first-order time derivatives with Eqs.\;\ref{eq4} and\;\ref{eq3} for the classical case) for both ionic species, which  
   leads to  the following equation for the density of charge: 
  \begin{equation}
_0D_t^{\alpha} \rho (x,t+\tau_1)  =   D_{\alpha} \frac{\partial^2 \rho(x,t+\tau_2)}{\partial x^2}  - D_{\alpha} \kappa^2 \rho (x,t+\tau_2)   
\label{eq27}
\end{equation}  
In detail this is done by first taking the divergence of the flux for both species (Eq.\;\ref{eqJa1} with separate delays), i.e.:
\begin{widetext}
\begin{equation}
\frac{\partial J_{+_{\alpha}} \left({x}, t + \tau_1 \right)}{ \partial x}   = - D_{\alpha} \frac{\partial }{ \partial x}\left\{
\left[ \frac{\partial }{ \partial x} + \frac{z F}{R T} \frac{\partial}{ \partial x} \phi \left({x}, t+ \tau_2 \right) \right] c_+ \left({x}, t + \tau_2 \right)  \right\}
\label{eqJp}
\end{equation}
\begin{equation}
\frac{\partial {J}_{-_{\alpha}} \left({x}, t + \tau_1 \right)}{ \partial x}   = - D_{\alpha} \frac{\partial }{ \partial x}\left\{
\left[ \frac{\partial }{ \partial x} - \frac{z F}{R T} \frac{\partial}{ \partial x} \phi \left({x}, t+ \tau_2 \right) \right] c_- \left({x}, t + \tau_2 \right)  \right\}
\label{eqJn}
\end{equation}
\end{widetext}
Subtracting Eq.\;\ref{eqJn} from \ref{eqJp},  multiplying by $z F$, and with the use of the fractional continuity equation and $2 c_0 = c_+ \left( {x}, t  \right) + c_- \left( {x}, t \right) $ we get:
\begin{widetext}
\begin{equation}
_0D_t^{\alpha} \rho \left( {x}, t + \tau_1\right) = D_{\alpha}   \left[ \frac{\partial^2 \rho \left({x}, t+ \tau_2 \right)}{ \partial x^2}
 + \frac{z^2 F^2}{R T} \frac{\partial }{ \partial x} \left[ 
 2c_0 
\frac{\partial \phi \left({x}, t + \tau_2 \right)}{ \partial x} \right] \right]
\end{equation} 
\end{widetext}
With the Poisson equation, we finally obtain the result given in Eq.\;\ref{eq27}. Eq.\;\ref{eq27} is 
similar to the classical PNP system \cite{BUCK1969219}:
\begin{equation}
\frac{\partial \rho (x,t)}{\partial t}  =   D \frac{\partial^2 \rho (x,t)}{\partial x^2}  - D \kappa^2 \rho (x,t)   
\label{eq6}
\end{equation} 
when $\tau_1=\tau_2=0$. 
It can also be rewritten  in dimensionless form as:
 \begin{equation}
 _0D_{\theta}^{\alpha} \rho (z,\theta+\theta_1)= \frac{\partial^2 \rho (z,\theta+\theta_2)}{\partial z^2}  -  \rho  (z,\theta+\theta_2)
%\label{eq8} 
\end{equation}
where $\theta = D_{\alpha} \kappa^2 t^{1/\alpha}$, $\theta_j = D_{\alpha} \kappa^2 \tau_j^{1/\alpha}$,  and $z= \kappa x$.

Now we apply the Laplace transform
(defined as $\tilde{f}(x,s) =\int_0^{\infty} f(x,t)e^{-st} \mathrm{d}t$) to both sides of Eq.\;\ref{eq27}, considering zero initial conditions, to obtain: 
  \begin{equation}
e^{-s \Delta \tau } s^{\alpha} \tilde{\rho} (x,s) = 
D_{\alpha} \frac{\mathrm{d}^2 \tilde{\rho} (x,s)}{\mathrm{d} x^2}  - D_{\alpha} \kappa^2 \tilde{\rho} (x,s)
\label{eq250}
\end{equation}
Rearranging gives:
\begin{equation}
\frac{\mathrm{d}^2 \tilde{\rho} (x,s)}{\mathrm{d}x^2} - \tilde{\eta}_{\alpha}^2\tilde{\rho } (x,s) = 0
\label{eq:29}
\end{equation}
 where
 \begin{equation}
\tilde{\eta}_{\alpha}^2 =  \frac{e^{-s \Delta\tau }s^{\alpha}}{D_{\alpha}} + \kappa^2  
\label{eq20}
\end{equation}
It is  straightforward to realize that when $\alpha=1$ and $\Delta\tau = \tau_2-\tau_1=0$, Eq.\;\ref{eq:29} reduces to the equivalent classical expression 
%\begin{equation}
 ${\mathrm{d}^2 \tilde{\rho}}/{\mathrm{d}x^2} -  \tilde{\eta}^2 \tilde{\rho} = 0
%\label{eq:12}
$ where $\tilde{\eta}^2 =  {s}/{D} + \kappa^2 $ \cite{gomez}. 
% \end{equation}
The general solution of Eq.\;\ref{eq:29} is of the form:
%\begin{equation}
$\tilde{\rho} (x,s) = \tilde{C_1} e^{-\tilde{\eta}_{\alpha} x} + \tilde{C_2} e^{\tilde{\eta}_{\alpha} x}$,  
%\end{equation} 
which with the   conditions $\rho({x},0) =0$,  
$\rho(\infty, t) =0$ and $\rho(0,t) =-  \kappa^2 \epsilon \phi_0(t)$ as used in Gomez-Zamudio et al. \cite{gomez}, we find:
\begin{equation}
\tilde{\rho} (x,s) = -\kappa^2 \epsilon \tilde{\phi}_0 e^{-\tilde{\eta}_{\alpha} x}
\label{eq10}
\end{equation}
Now the Laplace transform applied to the Poisson equation (Eq.\;\ref{mP})  with the result of Eq\;\ref{eq10} gives the following equation:
 \begin{equation}
\frac{\mathrm{d}^2 \tilde{\phi} (x,s)}{\mathrm{d}x^2} = \kappa^2   \tilde{\phi}_0 e^{-\tilde{\eta}_{\alpha} x}
\end{equation}
which, after integrating once and using the condition that $\tilde{i}(0,s)=0$,  {gives}:
\begin{equation}
{\frac{\mathrm{d} \tilde{\phi} (x,s)}{\mathrm{d}x} = - \frac{\kappa^2 \tilde{\phi}_0 e^{-\tilde{\eta}_{\alpha} x}}{\tilde{\eta}_{\alpha}}  + \frac{\tilde{\phi}_0(\kappa^2-\tilde{\eta}_{\alpha}^2)}{\tilde{\eta}_{\alpha}}}
\label{eq12}
\end{equation}
Using Eq.\;\ref{eq12} and the derivative of $\tilde{\rho} (x,s)$ (Eq.\;\ref{eq10}) w.r.t. $x$ we obtain the following expression for the Laplace transform of the current:
\begin{equation}
\tilde{i}_{\alpha} (x,s) =   \frac{ D_{\alpha}   \kappa^2 \epsilon \tilde{\phi}_0}{\tilde{\eta}_{\alpha}} \left( \kappa^2 - \tilde{\eta}_{\alpha}^2 \right) \left( e^{-\tilde{\eta}_{\alpha} x} - 1 \right)
%-D \frac{\text{d} \tilde{\rho}}{\text{d} x} - D   \kappa^2 \epsilon \frac{\text{d} \tilde{\phi}}{\text{d} x} 
\end{equation}
and thus a modified admittance function   as:
\begin{align}
Y_{\alpha} (x,s) &= Z_{\alpha}^{-1} (x,s)  = \frac{\tilde{i}_{\alpha} (x,s)}{\tilde{\phi}_0} \nonumber\\
 &=   \frac{ D_{\alpha}   \kappa^2 \epsilon }{\tilde{\eta}_{\alpha}} \left( \kappa^2 - \tilde{\eta}_{\alpha}^2 \right) \left( e^{-\tilde{\eta}_{\alpha} x} - 1 \right)
\label{eq32}
\end{align} 
The expression of $Y_{\alpha} (x,s)$ incorporates both the fractional order behavior of CPEs and a delay $\Delta\tau$ that limits  the system bandwidth at which energy storage takes place, as shall be shown in the next section. In principle, the measurement of admittance from voltage excitation and current response takes place at the boundaries of the EDLC system, and thus  $Y_{\alpha} (x,s) = Y_{\alpha} (x=x_h,s)$ where $x_h$ denotes for instance the thickness of the Helmholtz   layer.  This makes the admittance to be  solely dependent on the frequency in this case.

 To find the real and imaginary part of $Y_{\alpha} (x,s)$, we 
write 
 $\tilde{\eta}_{\alpha}^2 =  {e^{-s\Delta \tau }s^{\alpha}}/{D_{\alpha}} + \kappa^2  = a+ j b=  {(p+jq)^2}
$ where:
\begin{align}
a&=\frac{\omega^{\alpha} \cos(\alpha\pi/2 - \omega\Delta \tau )}{D_{\alpha}} + \kappa^2 \\
b&=\frac{\omega^{\alpha} \sin(\alpha\pi/2 - \omega\Delta \tau)}{D_{\alpha}}
\end{align}
which gives:
\begin{align}
p &= \frac{1}{\sqrt{2}} \sqrt{\sqrt{a^2+b^2} + a}\\
q&= \frac{\mathrm{sgn}\,{b}}{\sqrt{2}} \sqrt{\sqrt{a^2+b^2} - a}
\end{align}
where $\mathrm{sgn}\,{b}=b/|b|$. Thus:
\begin{align}
\frac{\mathrm{Re}(Y_{\alpha})}{D \kappa^2 \epsilon/(p^2+q^2)}
&= p(\kappa^2-p^2-q^2) (e^{-px}\cos(qx)-1) \nonumber\\
&-q(\kappa^2+p^2+q^2) e^{-px} \sin (qx) \\
\frac{\mathrm{Im}(Y_{\alpha})}{D \kappa^2 \epsilon/(p^2+q^2)}
&=- q(\kappa^2+p^2+q^2) (e^{-px}\cos(qx)-1) \nonumber\\
&-p(\kappa^2-p^2-q^2) e^{-px} \sin (qx)
\end{align}  
 from which the admittance phase angle is:
\begin{equation}
\phi(Y_{\alpha}) = \tan^{-1} [\mathrm{Im}(Y_{\alpha})/\mathrm{Re}(Y_{\alpha})]
\end{equation} 
 We can also express the results in terms of impedance such that:
 \begin{eqnarray}
&&{\mathrm{Re}(Z_{\alpha})} = {{\mathrm{Re}(Y_{\alpha})}}/{|Y_{\alpha}|^2}\\  
&&{\mathrm{Im}(Z_{\alpha})}  =-{{\mathrm{Im}(Y_{\alpha})}}/{|Y_{\alpha}|^2}\\ 
&&\phi(Z_{\alpha}) = - \phi(Y_{\alpha}) 
\end{eqnarray}
 which will be the focus of the simulation results below. 

\begin{figure*}[!t]
\begin{center}
\subfigure[]{\includegraphics[width=2.5in]{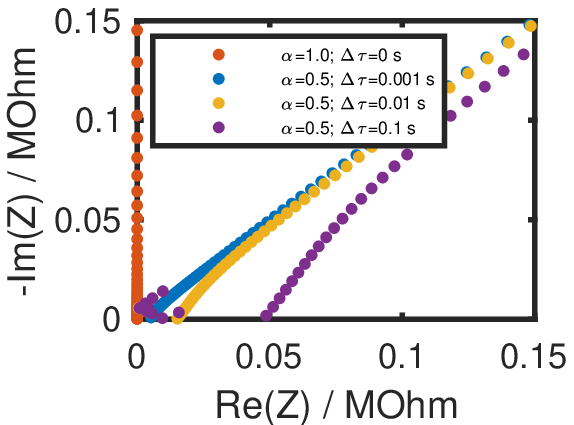}}
\subfigure[]{\includegraphics[width=2.5in]{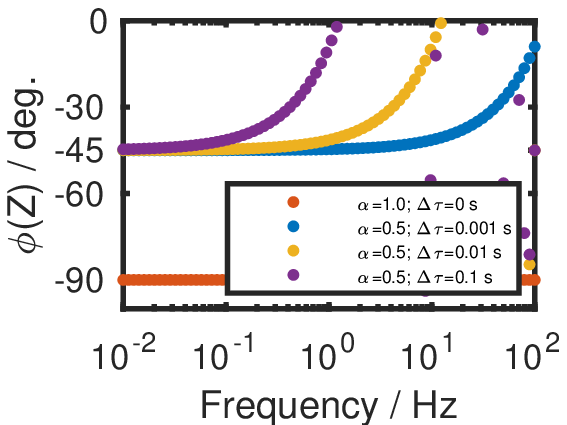}}
\subfigure[]{\includegraphics[width=2.5in]{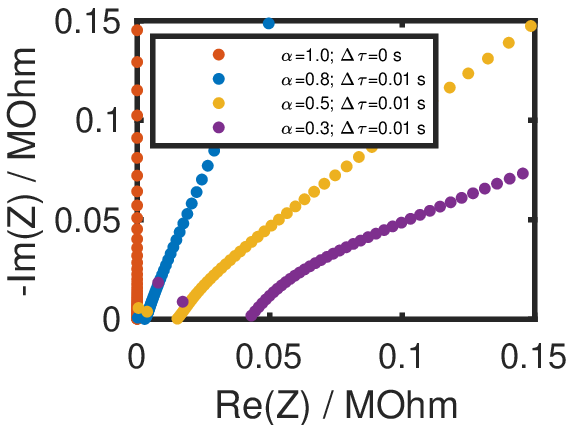}}
\subfigure[]{\includegraphics[width=2.5in]{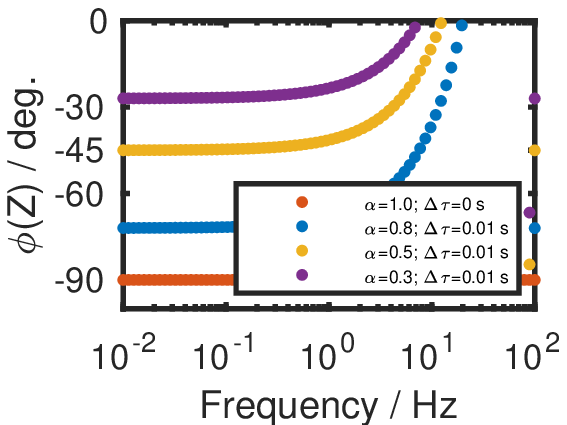}}
\caption{Plots of Eq.\;\ref{eq32} in Nyquist impedance representation and phase angle vs. log of frequency for (a)-(b) different value of $\Delta\tau$ and constant values of $\alpha=0.5$, and (c)-(d) different values of $\alpha$ and constant value of $\Delta\tau=0.01$\,s. The parameters are 
$x=\text{1\,nm}$; 
$T=\text{300\,K}$;
$D=2 \times 10^{-9}\,\text{m}^2\,\text{s}^{-1}$;
$z=\text{1}$;
$c_0=\text{1\,mmol\,L}^{-1}$;
$\epsilon=6.95\times 10^{-10}$ F\,m$^{-1}$, and 
the frequency points ($f=\omega/(2\pi)$) are logarithmically spaced from $10^{-2}$ to $10^{2}$ Hz with 20 pts per decade 
}
\label{fig1}
\end{center}
\end{figure*}

\section{Results and Discussion}
\label{simulation}

Plots of parametric imaginary vs. real parts of impedance  and  impedance phase angle vs. frequency using Eq.\;\ref{eq32} are given in Fig.\;\ref{fig1}. 
 The plots show the effects of $\Delta\tau$ varied between 0 and 0.1 s   (Figs.\;\ref{fig1}(a) and \ref{fig1}(b)) and the effect of $\alpha$ varied between 1.0 and 0.3, i.e. subdiffusive transport (Figs.\;\ref{fig1}(c) and \ref{fig1}(d)). 
We used the same parameters values as in  Gomez-Zamudio et al. \cite{gomez}, i.e.  
$T=\text{300\,K}$;
$D_{\alpha}=2 \times 10^{-9}\,\text{m}^2\,\text{s}^{-\alpha}$;
$z=\text{1}$;
$c_0=\text{1\,mmol\,L}^{-1}$;
$\epsilon=6.95\times 10^{-10}$ F\,m$^{-1}$, and 
 frequency points logarithmically spaced from $10^{-2}$ to $10^{2}$ Hz (20 pts per decade). 
 As for $x$ we set it equal to \text{1\,nm} which is   roughly the thickness of the Helmholtz layer, and at   which the current and thus the impedance are computed.

 As expected for the case of an ideal capacitor (i.e. $\Delta\tau=0$ and $\alpha=1$), the Nyquist plot is vertical with the high-frequency intercept on the real axis at practically zero Ohm. The impedance phase angle  is constant right at -90 deg. for all frequencies. Such  results are shown in all subfigures of Fig.\;\ref{fig1} as a comparative reference. 
 
 Now as we vary the value of the  time lag $\Delta \tau$ for a constant value of $\alpha=0.5$ (Warburg regime), it is clear from Fig.\ref{fig1}(b) that there is a critical frequency above which the impedance phase angle increases exponentially towards larger values than   -45 deg., observed at  low-frequencies. This is controlled by  the term  $e^{-s\Delta \tau  }$ in $\tilde{\eta}_{\alpha}$ that appears in the expression of the modified    impedance function, which, in the time domain differential equations is associated with $\Delta\tau = \tau_2-\tau_1$.  
It is important to mention that the values  of $\Delta \tau$ are positive in our case.
 This is different from standard Catteneo's formalism (purely due to diffusive transport, i.e. $\tau_2=0$) because of the fact that the current in the PNP system is the sum of both diffusion currents due to concentration gradients and drift currents due to potential gradient.  
 As a consequence,  
 the net value of  $\Delta\tau$  
 can be   thought of as the resulting time shift  due to both   diffusion  and drift. For the diffusion transport, it is the variation of concentration that leads to the diffusion current after a certain time delay, 
whereas  the drift currents, from both positive and  negative charges because of the electric field, lead to    changes  in the concentration of the different species after    another time delay. 
 It is clear that with the  positive values of $\Delta\tau$,  the  results are very comparable to real  resistive-capacitive devices as shown in Fig.\;\ref{fig0} in which (i) both real and imaginary parts of the impedance are frequency-dependent, but (ii) also is the impedance phase angle especially at intermediate frequencies away from dc. This is not case for a single CPE that is by definition known to exhibit a constant impedance phase angle at all frequencies.  
 
 As for the effect of the fractional coefficient $\alpha$ for a fixed value of $\Delta \tau$ (taken as 0.01 s) on the spectral impedance,  Figs.\ref{fig1}(c)   and\;\ref{fig1}(d) show clear angular  deviation of the Nyquist plot and phase angle plot from that of ideal capacitors  as the values of $\alpha$ are decreased further away from one. At the low frequencies limits, $\phi(Z) \to -\alpha \pi/2$ as expected for a CPE.  
  With $\alpha=1$, one recovers the normal electrodiffusion model and thus the pure capacitive behavior.   
 Again the presence of non-zero value for $\Delta\tau$ in the expression of the modified impedance makes the system  veer towards a resistor   as the frequency is increased.

In summary, the anomalous  electrodiffusion problem described with fractional-order calculus is known to give  rise to  the  CPE  behavior,  which  appears as a straight line in the Nyquist impedance representation, or a constant impedance phase for all frequencies. However, this is not the case for experimental results of  EDLCs for instance, which show a CPE behavior only for a limited bandwidth at close-to-dc frequencies and a tendency towards resistive behavior as the frequency is increased (Fig.\;\ref{fig0}). 
Lenzi et al. have just recently demonstrated that with the use of fractional   differential operators with non-singular kernels (i.e., Atangana-Baleanu \cite{atangana2016new}) it is possible the retrieve the impedance response of a CPE in  association with a resistor, which is in very good agreement with the experiment. Nevertheless, this approach relies on properly choosing a mathematical expression for the kernel used in the fractional integro-differential equation, such as the generalized two-parameter Mittag-Leffler function for instance, with which the physical  traceability  may be blurred.     
Our modified PNP scheme   with a time shift between cause and effect, and with mono-order fractional time derivatives also successfully captured such a resistive-capacitive behavior in a compact and concise way. While the physical meaning of the fractional exponent is still under debate, the use of delayed equations is believed to be more  advantageous to explain the physical mechanism in fractional EDLCs \cite{foedlc}. 

\section{Conclusion}
\label{conclusion}

In the standard electrokinetic PNP model, the Poisson equation relates the free charge density to the Laplacian of the electric potential, and the transport of dilute ionic species is governed by the fluxes of electromigration and diffusion. The continuity equation connects the first-order time derivative of ion concentrations to the divergence of the the total flux. 
Here we  proposed  an extended fractional  time derivative of  time-shifted concentrations of electrodiffusing species as a modified model for the PNP system. 
 The generalization of  the integer-order time derivative to  fractional-order time derivative  is   commonly employed for describing the behavior of complex system when integer-order models fail. 
Considering blocking electrodes conditions, the model provided close-to-realistic description of typical EDLC impedance profile at low-to-intermediate frequencies. 
 With  fractional exponents different from one, the impedance phase angle at low frequencies is shown to deviate from the  response of ideal capacitors (i.e. CPE impedance), whereas non-zero values of the   time shift $\Delta\tau$ allowed the modeling of the increased resistive behavior as the frequency is increased.
 The  time shift is due to the resulting net effect of both diffusion and drift time shifts with respect to changes in concentration of diffusing species. 
  In principle, other boundary conditions such as Faradaic currents at the electrodes can be investigated for simulating the impedance profiles of electrochemical  sensors or batteries.

\section*{Acknowledgement}

This work was supported by the NSF project \#2126190 (C.W \& A.A.)

\section*{Data Availability} 

The data that supports the findings of this study are available within the article.

\bigskip

%\bibliography{biblio}

%merlin.mbs aipnum4-1.bst 2010-07-25 4.21a (PWD, AO, DPC) hacked
%Control: key (0)
%Control: author (8) initials jnrlst
%Control: editor formatted (1) identically to author
%Control: production of article title (0) allowed
%Control: page (1) range
%Control: year (1) truncated
%Control: production of eprint (0) enabled
%

 \end{document}